\documentclass[prl,nofootinbib,twocolumn,superscriptaddress]{revtex4}

\usepackage{graphicx} \usepackage{bm} \usepackage{epsfig}
\usepackage{amsmath}

\newcommand{\beq}{\begin{equation}}
\newcommand{\eeq}{\end{equation}}
\newcommand{\bea}{\begin{eqnarray}}
\newcommand{\eea}{\end{eqnarray}}

\begin{document}
\title{Two component dark matter}

\author{Malcolm Fairbairn}
\affiliation{Physics, King's College London, Strand, London WC2R 2LS}
\affiliation{Theory Division, Department of Physics, CERN, CH-1211 Geneva 23,
Switzerland}

\author{Jure Zupan}
\affiliation{Theory Division, Department of Physics, CERN, CH-1211 Geneva 23,
Switzerland}
\affiliation{J.~Stefan Institute, Jamova 39, 1000 Ljubljana, Slovenia}
\affiliation{Faculty of mathematics and physics, University of Ljubljana, Jadranska 19, 1000 Ljubljana, Slovenia}

\begin{abstract}
We explain the PAMELA positron excess and the PPB-BETS/ATIC $e^++e^-$ data using a simple two component dark matter model (2DM). The two particle species in the dark matter sector are assumed to be in thermal equilibrium in the early universe. While one particle is stable and is the present day dark matter, the second one is 
metastable and decays after the universe is $10^{-8}$ s old. In this model it is simple to accommodate the large boost factors required to explain the PAMELA positron excess 
without the need for large spikes in the local dark matter density. We provide the constraints on the parameters of the model and comment on possible signals at future colliders. 
\end{abstract}

\maketitle

\section{Introduction}
Other than its gravitational interactions, the properties of dark matter are largely unknown - only lower limits on the mass and upper limits on the couplings have so far been obtained.  It has been suggested that the preliminary results from The Payload for Antimatter Matter Exploration and Light-nuclei Astrophysics (PAMELA) experiment may represent a breakthrough in this situation \cite{Adriani:2008zr}.  PAMELA sees a larger positron fraction in the cosmic ray flux at 10 - 80 GeV than one expects in the galactic environment, in agreement with previous hints from the HEAT and AMS-01 experiments \cite{Barwick:1997ig,Aguilar:2007yf}. A possible explanation for this excess is that it comes from dark matter (DM) annihilating in our own galaxy \cite{Chen:2008dh,Cirelli:2008pk,Cholis:2008hb,Finkbeiner:2008qu,Rothstein:2009pm,Harnik:2008uu,Nomura:2008ru,Fox:2008kb,Chen:2009dm,Mardon:2009rc,Meade:2009rb,Grajek:2008pg}
or from DM decaying with a decay time much longer than the age of the universe 
\cite{Bertone:2007aw, Nardi:2008ix, Ibarra:2008jk,Arvanitaki:2008hq,Ishiwata:2009vx,Shirai:2009kh,Cheung:2009si} (or the combination of the two \cite{Cheung:2009si}).
Astrophysical sources, such as a nearby pulsar, have also been suggested as an explanation of the excess \cite{taylor,backer,Hooper:2008kg, Yuksel:2008rf, Profumo:2008ms, Shaviv:2009bu,Malyshev:2009tw, Hall:2008qu}. In this paper we will focus on the annihilating DM interpretation of the excess.

The PAMELA experiment also detects an anti-proton cosmic ray spectrum which  is compatible with the expected galactic background, giving a constraint on dark matter annihilation modes.  The authors of Ref. \cite{Cirelli:2008pk} find two explanations for the PAMELA data: {\it i}) a heavy DM particle of mass above about 10 TeV that annihilates predominantly to a $W^+W^-$ pair or {\it ii}) a DM particle which annihilates predominantly into SM leptons with no strong constraint on the DM mass.
  
There are also hints of a positron excess in the energy range of $500-800$ GeV from a long duration Polar Patrol Balloon (PPB-BETS) flight and from the ATIC-2 balloon experiment \cite{Torii:2008xu,ATIC-2}, in agreement with the change of power-law seen by HESS above $\sim 1$ TeV~\cite{Collaboration:2008aaa}. If one tries to explain simultaneously both these data and the PAMELA data using DM annihilations, it seems necessary to consider a DM particle with mass around 1 TeV which annihilates into leptons \cite{Chen:2008dh,Cirelli:2008pk,Finkbeiner:2008qu,Rothstein:2009pm}.

One class of DM candidates are particles that have a weak-scale self-annihilation cross section at freeze-out \cite{Bertone:2004pz} 
\beq\label{sigmaCDM}
\langle \sigma_A v\rangle_{\rm F} \simeq 3 \times 10^{-26} {\rm cm}^3/{\rm s}. 
\eeq
The resulting thermal relic abundance is then close to what is required to explain dark matter. Such a candidate does not, however, fit straightforwardly with the PAMELA observations. 
If the PAMELA excess flux of positrons, $\Phi_{e^+}$,
is due to DM annihilating in the milky way, then 
\beq\label{Phi_propto}
\Phi_{e^+}\propto \bar N_{e^+} \langle \sigma_A v\rangle \rho_{\rm DM}^2/m_{\rm DM}^2,
\eeq
where $\rho_{\rm DM}$ is the local DM density, $m_{\rm DM}$ the DM particle mass and $\bar N_{e^+}$ the average number of positrons produced in a single $\chi\chi\to X$ annihilation (usually $\bar N_{e+}\leq 1$ and at most a factor of few in more exotic models). The proportionality in \eqref{Phi_propto} includes among others the effect of positron propagation in galactic medium. For masses above $100$ GeV the PAMELA positron excess suggests a $\langle \sigma_A v\rangle$ which is larger than the value required at freeze-out $\langle \sigma_A v\rangle_{\rm F}$ \cite{Cirelli:2008pk,Cholis:2008hb,Finkbeiner:2008qu,Rothstein:2009pm}.  The mismatch is parametrized by a parameter called the boost factor
\beq\label{boost-factor}
B\equiv \bar N_{e^+} \frac{\langle \sigma_A v\rangle\,\rho_{\rm DM}^2}{\langle \sigma_A v\rangle_{F}\,(\bar \rho_{\rm DM})^2},
\eeq
where $\bar \rho_{\rm DM}\simeq0.35~{\rm GeV}/{\rm cm}^3$ is the average expected local DM density. 
We can have $B>1$ if either $\langle \sigma_A v\rangle>\langle \sigma_A v\rangle_{F}$ or if there is a local DM over-density due to substructure in the galactic distribution\footnote{Conventionally, "boost factor" refers only to this astrophysical origin of the cosmic ray enhancement, while we prefer to define it as referring to the combined effect.}.  Analysis of the N-body dark matter only Via Lactea II simulation suggests there is a $1\%$ probability of density fluctuations enhancing the local annihilation rate by a factor $B\sim 10$, while much smaller fluctuations were observed in the Aquarius Project simulations \cite{Diemand:2008in,Springel:2008by}. While substructure in simulations is not well understood, larger boost factors seem to require a different explanation -- 
an enhanced DM annihilation cross section. The values of $B$ which explain the positron data are roughly $B\sim (10,10^3, 10^5)$ for $m_{\rm DM}\sim (100 {\rm ~GeV}, 1 {\rm ~TeV}, 10 {\rm ~TeV})$ respectively, with a variation of a factor of few depending on the predominant annihilation channel \cite{Cirelli:2008pk,Cholis:2008hb}.  

If the annihilation proceeds through the $s-$channel, then $\sigma_A\propto 1/v$ and $\langle \sigma_A v\rangle_{\rm F}$ at freeze-out is $\simeq \langle \sigma_A v\rangle$ today, 
giving no enhancement in \eqref{boost-factor}.  If the annihilation proceeds through the $p-$channel then $\sigma_A\propto v$ meaning that $\langle \sigma_A v\rangle\ll \langle \sigma_A v\rangle_{\rm F}$.  In both cases the DM interpretation of the PAMELA data  for
 $m_{\rm DM}>100$ GeV seems to exclude a simple single component thermal relic explanation of the results. This is true also, if DM is composed of many stable components as in \cite{Cao:2007fy}, in which case $B<1$. In the case of co-annihilation, the boost factor is bounded by $B<N_f^2$, where $N_f$ is the number of co-annihilating flavors. Again, large boost factors are excluded unless $N_f$ is very large.

If, however, the $s-$channel annihilation is enhanced by Sommerfeld corrections, then $\langle \sigma_A v\rangle \gg \langle \sigma_A v\rangle_{\rm F} $ is possible. For DM particles which interact with $W,Z$ bosons, these Sommerfeld corrections are present if the DM particles are heavy, $m_{DM}\gtrsim 4\pi m_W/g^2\sim 2$ TeV. The annihilation cross section is then enhanced by $g^2/v$ and can become large for $v\to 0$ \cite{Hisano:2003ec}. Alternatively, the Sommerfeld enhancement can come from a new force in the DM sector where a force carrier possesses GeV mass \cite{Finkbeiner:2008qu}, with very large Sommerfeld enhancements possible,
if there exists a bound state very close to threshold \cite{MarchRussell:2008tu,Pospelov:2007mp}. Enhancements are also possible if the annihilation goes through a resonance with mass of order $2m_\chi$ \cite{Ibe:2008ye,Guo:2009aj}.  

All these explanations require light new degrees of freedom with mass below $m_{DM}$. We will address the other possibility, namely 
that the apparent enhancement comes from states that are heavier than $m_{DM}$. As a working tool we will present a simple 2-component dark matter (2DM) model that can explain the PAMELA and balloon experiment data \footnote{The name is somewhat of a misnomer since only one of the two components is stable on cosmological timescales.}.

\section{Two component dark matter}
In the 2DM model the dark sector is composed of two particle flavours. These two DM particles can be scalars, fermions or vectors.
The first DM particle $\chi_1$ has mass $m_1$ and is stable. We assume that it is this particle which is the cold dark matter (CDM) relic 
still present today, 
responsible for galactic rotation curves and also, through self-annihilation, the PAMELA and balloon results. The second DM particle $\chi_2$ has mass $m_2>m_1$, is unstable and decays to $\chi_1$. We also assume that both $\chi_2$ and $\chi_1$ are in thermal equilibrium
prior to freeze out. We will discuss under what conditions such a setup is able to explain PAMELA and balloon data.
Many decay modes of $\chi_2$ are possible including a two body decay $\chi_2\to 2 \chi_1$ or multibody decays $\chi_2\to \chi_1+X$, where $X$ are SM particles. To retain generality we define $N_{\rm dec}$ as the average number of $\chi_1$ particles produced in a single $\chi_2$ decay.

The number density of each of the dark matter flavours evolves with time as follows
\beq
\begin{split}
\frac{d n_1}{dt} +3 H n_1&=-\langle \sigma_{A1} v_1\rangle \big(n_1^2-n_{1{\rm eq}}^{2}\big)+N_{\rm dec} \Gamma_2 n_2,\\
\frac{d n_2}{dt} +3 H n_2&=-\langle \sigma_{A2} v_2\rangle \big(n_2^2-n_{2{\rm eq}}^{2}\big)-\Gamma_2 n_2,
\end{split}
\eeq
where $n_{i{\rm eq}}(t)$ is the equilibrium number density given by the thermal Boltzmann distribution and $\Gamma_2$ is the decay width for $\chi_2\to N_{\rm dec} \chi_1+X$ decay. It is useful to define $z=m_1/T$ and normalise number density to entropy density, $Y_i(z)=n_i(z)/s(z)$. Then the evolution equations become
\beq\label{Yi evolution}
\begin{split}
\frac{z}{Y_{1{\rm eq}}}\frac{dY_1}{dz}&=-\frac{\Gamma_{A1}}{H}\Big[\Big(\frac{Y_1}{Y_{1{\rm eq}}}\Big)^2-1\Big]+N_{\rm dec} \frac{\Gamma_2}{H} \frac{Y_2}{Y_{1{\rm eq}}},\\
\frac{z}{Y_{2{\rm eq}}}\frac{dY_2}{dz}&=-\frac{\Gamma_{A2}}{H}\Big[\Big(\frac{Y_2}{Y_{2{\rm eq}}}\Big)^2-1\Big]- \frac{\Gamma_2}{H} \frac{Y_2}{Y_{2{\rm eq}}},
\end{split}
\eeq
where the annihilation rates are $\Gamma_{Ai}=n_{i{\rm eq}}\langle \sigma_{Ai} v_i\rangle$, while $Y_{i{\rm eq}}(z)=n_{i{\rm eq}}(z)/s(z)$.

\begin{figure}
\includegraphics[width=7.2cm]{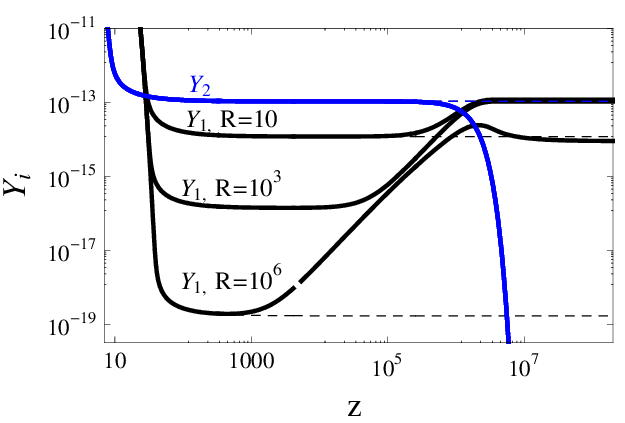}
\includegraphics[width=7.2cm]{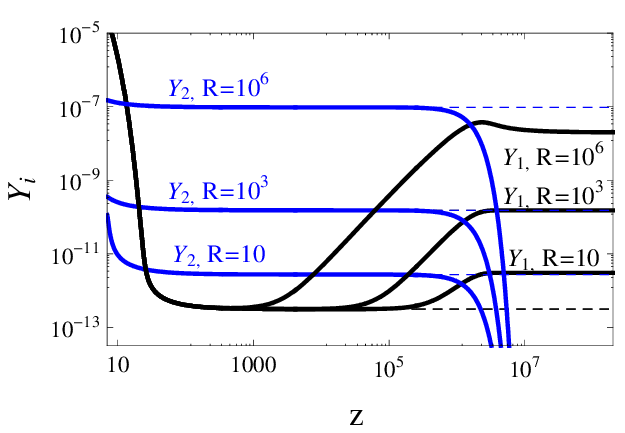}
\vskip-0.3cm
\caption[1]{The solution for $Y_1(z)$ (black solid line) and $Y_2(z)$ (blue solid line) for $m_1=1$ TeV, $m_2=3$ TeV,  $\Gamma_2=10^{-24}$ GeV, $N_{\rm dec}=1$ and three different values of $R$ as denoted.  Dashed lines denote the thermal relic values of $Y_i(z)$. On the upper figure $\langle \sigma_{A2} v_2\rangle$ is held fixed to $\langle \sigma_A v\rangle_{\rm CDM}=3\times 10^{-26} {\rm cm}^3/{\rm s}$, so that without decay this would give correct DM relic density with $\chi_2$ the DM particle. Through decay this is transfered to $\chi_1$. For illustration we also show the lower figure, where $\langle \sigma_{A1} v_1\rangle$ is held fixed to $3\times 10^{-26} {\rm cm}^3/{\rm s}$. }
\label{YiPlot} 
\vskip-0.5cm
\end{figure}

An interesting limit to consider is $\Gamma_2\ll \Gamma_{A1,2}$. The abundances then first settle into their thermal relic values, $Y_i\to Y_i^{\rm Th.rel.}$, after which $\chi_2$ decays to $\chi_1$ (see Fig. \ref{YiPlot}). In this limit we then have
\beq
Y_1(\infty) =Y_1^{\rm Th. rel.}+N_{\rm dec}Y_2^{\rm Th. rel.}.
\eeq
The number density of the dark matter relic is enhanced by the number density of metastable dark matter components. 
Using an approximate analytic solution for s-wave annihilation, where  $Y_i^{\rm Th. rel.}\propto 1/\langle \sigma_{Ai} v_i\rangle$, we find 
\beq\label{Y1result}
 \frac{Y_1(\infty)}{Y_1^{\rm Th. rel.}}\simeq 1+N_{\rm dec} R\Big (1-\frac{1}{z_1^F}\log R\Big), 
 \eeq
 where $z_1^F\simeq 20 $ is the freeze out value of $z$ for the stable DM component, while 
 \beq
R=\frac{m_1}{m_2} \frac{\langle \sigma_{A1} v_1\rangle }{\langle \sigma_{A2} v_2\rangle }.
\eeq

We can now see why it is possible to explain the PAMELA and balloon data using the 2DM model.   
If $R>1$ we will have  $ Y_1(\infty) > Y_1^{\rm Th. rel.}$.  The positron flux excess measured by PAMELA and given by $Y_1(\infty)^2 \langle \sigma_{A1} v_1\rangle$ will therefore be larger than one would expect in the case of one-component DM model, where the positron excess is proportional to $\big(Y_1^{\rm Th. rel.}\big)^2 \langle \sigma_{A1} v_1\rangle$. This is shown on Fig. \ref{YiPlot}. The larger the ratio $R$, the larger the enhancement of $Y_1(\infty)$. Note also that as the value of $\Gamma_2$ approaches the values of $\Gamma_{A1,A2}$ there is a washout effect, as seen for the $R=10^6$ curve in Fig. \ref{YiPlot}. If $\chi_2$ decays too quickly, then the resulting $\chi_1$ particles may still have chance to annihilate with each other before $\chi_1$ freezes out completely.  If we are not therefore in the limit where $\Gamma_2\ll \Gamma_{A1,A2}$, the enhancement effect is lost.

\begin{figure}
\includegraphics[width=6.2cm]{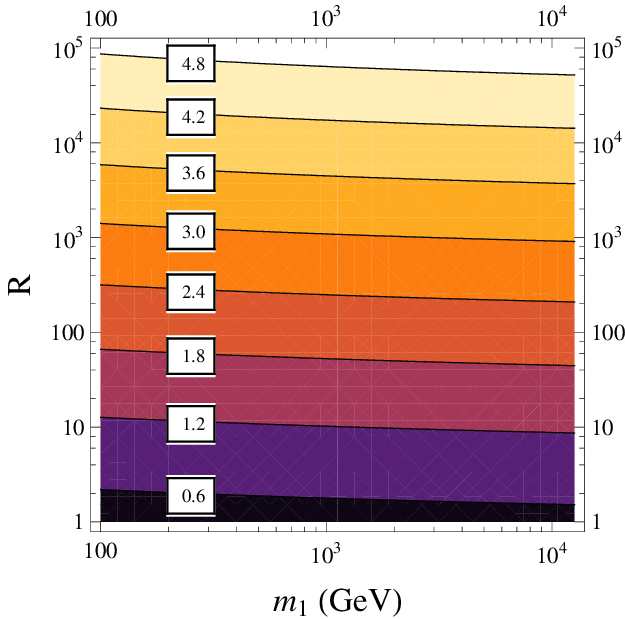}
\vskip-0.3cm
\caption[1]{The boost factor $B$ as a function of $m_1$ and $R$, in the case where $\Gamma_2$ is small enough that wash-out can be neglected. The framed numbers ($n$) labeling contours give boost factors as $B=10^n$. }
\label{Rplot} 
\vskip-0.5cm
\end{figure}

We also show in Fig. \eqref{Rplot} the predicted boost factors in the limit of small $\Gamma_2$ 
\beq\label{BF}
B=\frac{\langle \sigma_{A1} v_1\rangle}{\langle \sigma_{A} v\rangle_{\rm F}}\simeq\frac{z_{1F}^{\rm 2DM}}{z_F^{\rm CDM}}\sqrt{\frac{g_*^{\rm CDM}}{g_*^{\rm 2DM}}}  \frac{Y_1(\infty)}{Y_1^{\rm Th. rel.}}.
\eeq
The first two ratios on the right-hand-side are $O(1)$, while the last ratio is given in \eqref{Y1result}. The quantities labeled ${\rm CDM}$ are the parameters for the usual WIMP cold dark matter scenario and ${\rm 2DM}$ corresponds to the parameters of our model.  

We next discuss the limits on the parameters of our model. In order to explain the PAMELA data one needs $m_1$ to be larger than $50-100$ GeV, with $m_1\sim 1$ TeV suggested by the ATIC excess. For $\chi_2$ to be able to decay into $\chi_1$ we need  $m_2> m_1$. For a particular assumed annihilation channel $\chi_1\chi_1\to X_{\rm SM}$, the PAMELA and balloon experiment data fix $\langle \sigma_{A1} v_1\rangle=B(m_1)\langle \sigma_{A} v\rangle_F$ as a function of $m_1$ \cite{Cirelli:2008pk}. The value of $m_2\langle \sigma_{A2} v_2\rangle$ then follows from Fig. \ref{Rplot}. For large boost factors we have $\langle \sigma_{A2} v_2\rangle \ll \langle \sigma_{A1} v_1\rangle$.

To have significant boost factors, the washout effect should be small, giving an upper limit on $\Gamma_2$.  This means that at freeze-out, $z_1^F$, the second term in \eqref{Yi evolution} is smaller than the first one (we approximate $Y_{2}(z)$ at late times with its freeze-out value)
\beq\label{Gamma2}
\Gamma_2 \ll \frac{\Gamma_{A1}(z_1^F)}{ N_{\rm dec}} \frac{Y_{1{\rm eq}}(z_1^F)}{Y_{2{\rm eq}}(z_2^F)}\sim \frac{\Gamma_{A1}(z_1^F)}{ B},
\eeq 
and the annihilation rate at freeze-out is approximately
\beq\label{GammaA1}
\Gamma_{A1}(z_1^F)\sim 3H(z_1^F)\simeq
0.1 \frac{m_1^2}{m_{Pl}},
\eeq
where in the last equality we used $z_1^F\simeq 20$ and $g_*\simeq 10^2$.  
Equations \eqref{Gamma2} and \eqref{GammaA1} imply an upper bound on $\Gamma_2$
\beq\label{Gamma2bound}
\Gamma_2\ll 10^{-17} {\rm GeV}\cdot \Big(\frac{10^3}{B}\Big)\cdot \Big(\frac{m_1}{1{\rm TeV}}\Big)^2,
\eeq 
or 
\beq
\tau_2 \gg (10^{-7} s)\times \Big(\frac{B}{10^3}\Big)\cdot \Big(\frac{1{\rm TeV}}{m_1}\Big)^2.
\eeq 
Because of this, if $\chi_2$ is produced in colliders at relativistic velocities, it will travel at least a few meters before decaying.

There also exists a lower bound on $\Gamma_2$ from nucleosynthesis.  One can use the detailed results presented in \cite{kawasaki} to argue that so long as the lifetime of $\chi_2\to \chi_1+SM$ is less than around a second there will be no change to light element abundances.  
If on the other hand only $\chi_2\to \chi_1\chi_1$ is allowed, this usually does not affect abundances. The $\chi_2$ decay time can then be very long, of order the cosmological time scale. The simplest scenario then is that by the time of structure formation the $\chi_1$ particles from the decay are non-relativistic so that they are cold dark matter.   

More interesting is the situation where the particles decay after the start of structure formation.  In the 1980s a significant amount of thought went into the idea of decaying dark matter and its effect on structure formation.  The motivation at that time was the idea that the missing energy in the universe now usually subscribed to being dark energy was actually the relativistic decay products of dark matter \cite{turner, gelmini, dicus}.  Such particles would not cluster below 100 Mpc and therefore would not contribute to the value of $\Omega_M\ll 1$ measured on scales smaller than 100 Mpc.  This scenario is disfavoured by more recent observations of H(z) and structure formation and would not anyway solve the problem of the excess positrons we are addressing in this work.  However, work done at the time and more recently has shown that the decay of dark matter into another dark matter species (cold or hot) and its heating in that decay has interesting effects on structure, puffing out dark halos making them more diffuse \cite{khlopov, olive}.  Because of this some workers have suggested that decaying dark matter could have a bearing upon two apparent possible problems with the $\Lambda CDM$ scenario, namely the cuspy halo problem and the small scale power problem \cite{sanchez-saledo, cen,melia}.  While these problems are controversial in that not everybody actually agrees if they exist, it is certainly true that the possibility of dark matter decaying into another dark species has a rich phenomenology when it comes to structure formation.

\section{Particle physics context}
To recapitulate, we have found that the 2DM model with two DM components, $\chi_1$ that is stable, and $\chi_2$ that is metastable, can explain the enhanced annihilation cross section observed by PAMELA. The large boost factors observed are explained by the hierarchy 
\beq
\frac{m_1}{m_2} \frac{\langle \sigma_{A1} v_1\rangle }{\langle \sigma_{A2} v_2\rangle }\sim \frac{m_1}{m_2} \frac{g_1^4}{g_2^4}\frac{\Lambda_2^2}{\Lambda_1^2}\gg 1,
\eeq
where we have denoted schematically the dependence of cross sections in the non-relativistic limit on couplings $g_{1,2}$ and masses $\Lambda_{1,2}$ of exchanged particles. The hierarchy of annihilation cross sections $\langle \sigma_{A1} v_1\rangle \gg \langle \sigma_{A2} v_2\rangle $ can be obtained for instance, if the typical coupling $g_1$ in the first DM sector is larger than $g_2$ of the second one (here the hierarchy need not to be very large, for instance even for a boost factor of $10^4$, $g_2 \sim0.1 g_1$ suffices).
The other possibility is that the annihilation of $\chi_1$ proceeds through a heavier state than the annihilation of $\chi_2$. For large boost factors a relatively large hierarchy is needed, though (for instance for a boost factor of  $10^4$, $\Lambda_1 \sim10^{-2} \Lambda_2$). Another elegant possibility is that the annihilation of $\chi_1$ proceeds through and s-wave process, while the annihilation of $\chi_2$ is p-wave suppressed. The annihilation cross section for $\chi_2$ is then $v^2$ suppressed (see e.g. \cite{Goldberg:1983nd}), which for $v\sim 0.05$ at freeze out can lead to a boost factor of $\sim 10^3$ without any fine-tunings. This is easily realized in a concrete model, if $\chi_1$ is a Dirac fermion, while $\chi_2$ is a Majorana fermion. Yet another possibility is, if $m_2$ annihilation is phase space suppressed. For instance, if $m_2$ and $m_1$ are almost mass degenerate and $\chi_2 \chi_2\to \chi_1\chi_1$ is the dominant annihilation channel for $\chi_2$, while $\chi_1$ can also annihilate to SM. 

More severe is the hierarchy between $\Gamma_2$, Eq. \eqref{Gamma2bound}, and the decay width $\Gamma  \sim g^2 m/(16\pi)\sim $ few GeV typical for a weakly coupled theory. One explanation would be that $\chi_2$ carries an approximately conserved charge that suppresses its decay. A simple possibility is that $\chi_1$ is charged under $Z_2$ and $\chi_2$ under a different $Z_2'$, while the SM is neutral under $Z_2\times Z_2'$.
Then the annihilations $\chi_1 \chi_1\to X_{SM}$, $\chi_2\chi_2\to X$ are allowed, while $\chi_{1,2}\to X$ decays are not. If $Z_2'$ is broken at some high scale $\Lambda\gg 1$ TeV,  $\chi_2$ is metastable in agreement with the 2DM explanation of the PAMELA data. Another possibility which we do not pursue here is that $\chi_2$ could be charged under a gauge group broken at a high scale.

\begin{table}[t]
\begin{tabular}{c|rrr}
\hline\hline
 field & $\chi_1$ & $\chi_2$ & SM  \\\hline
 $Z_2$ & $-1$ & $1$& $1$\\
 $Z_2'$ & $1$ & $-1$& $1$\\\hline\hline
\end{tabular}
\caption{The charges under $Z_2\times Z_2'$.}\label{tab:charges}
\end{table}

We next discuss the possible interactions $\chi_1$ and $\chi_2$ can have. For simplicity we focus on the case, where both $\chi_1$ and  $\chi_2$ are scalars.
If $\chi_{1,2}$ are singlets under $SU(2)_L\times U(1)_Y$, then the most general renormalizable Lagrangian invariant under $Z_2\times Z_2'$ is 
\beq\label{Lchi}
\begin{split}
{\cal L}_{\chi}=&{\cal L}_{\rm kin}+ c_1 \chi_1^2\chi_2^2+c_2(H^\dagger H)\chi_1^2\\
&+c_3(H^\dagger H)\chi_2^2+c_4 \chi_1^4+c_5\chi_2^4.
\end{split}
\eeq
The $\chi_1$ and $\chi_2$ thermalise through interactions with the SM higgs and through four scalar interactions $c_{1,4,5}$ in the dark sector. In the milky way $\chi_1$ annihilate into $hh, WW, ZZ$ leading to leptonic and hadronic final states.  Thus $\chi_1$ mass has to be large enough, $m_1\sim 10 $ TeV so that $\chi_1\chi_1\to W^+W^-$ can explain both the $e^+/(e^++e^-)$ excess as well as the absence of $\bar p/p$ signal by PAMELA (while the balloon experiments, if confirmed, would exclude this simple scenario) \cite{Cirelli:2008pk}. It is possible to avoid this constraint by enlarging the dark sector, for instance by charging $\chi_1$ under an extra U(1) under which also the SM leptons are charged but not the quarks, while keeping $c_2$ small enough. 

To get large boost factors in this simple scenario the hierarchy $|c_2|\gg |c_{1,3}|$ is needed, for instance, for $B\sim 10^4$ a hierarchy $|c_2|\lesssim 10^{-2} |c_{1,3}|$ works. Also, in order to prevent ``fast" $\chi_2$ decay, the $Z_2'$ should not be broken by a $\chi_2$ vev (i.e. $\langle \chi_2\rangle=0$, which for instance is trivially true if all $c_i>0$).  These qualitative conclusions do not change even if $\chi_1$ and/or $\chi_2$ are neutral components of some higher representation of $SU(2)_L\times U(1)_Y$.  Even then the terms invariant under $Z_2\times Z_2'$ will contain two $\chi_1$ or two $\chi_2$ fields. Renormalizable interactions can thus couple $\chi_{1,2}$ only to the higgs or to $W,Z$. 

The dimension 5 operators relevant to the $\chi_2$ decay that break $Z_2'$ but not $Z_2$ are
\beq\label{dim5}
\begin{split}
&\frac{1}{\Lambda}\big(HH^\dagger\big)\chi_1^2\chi_2, \quad \frac{1}{\Lambda}\partial_\mu\chi_1\partial^\mu \chi_1 \chi_2,\quad \frac{m^2}{\Lambda} \chi_1^2 \chi_2,\\
&\frac{m}{\Lambda}\big(\bar \psi \{1,\gamma_5\} \psi\big)\chi_2, \quad\frac{m^2}{\Lambda}\big(H^\dagger H)\chi_2, \quad\frac{1}{\Lambda}\big(H^\dagger H)^2\chi_2,
\end{split}
\eeq
where $\psi$ are the SM fermions. The dimensionful parameter $m$ is most conveniently chosen to be $\sim m_2$ (the choice of $m$ only rescales the definition of $\Lambda$).  Decays into all massive standard model particles are possible with partial decay widths 
\beq \label{Gamma2explicit}
\Gamma_2\simeq m_2^3/(16 \pi \Lambda^2 ).
\eeq
The last and the first operator in \eqref{dim5} give decay widths that are additionally $(v/m_2)^4$ suppressed. From Eq. \eqref{Gamma2bound} and \eqref{Gamma2explicit} we then have (taking $m_1\sim m_2\equiv m_{1,2}$ for simplicity)
\beq\label{Lambdabound}
\Lambda\gg 10^{12}{\rm GeV}\cdot \Big(\frac{B}{10^3}\Big)^{1/2}\Big(\frac{m_{1,2}}{1{\rm TeV}}\Big)^{1/2}.
\eeq
This general limit applies also, if $\chi_{1,2}$ are fermions or vectors.

If we assume that $\chi_2$ decays with $\tau_2<1 s$ (so that BBN constraints are fulfilled regardless of the decay mode), this then implies
an upper limit on $\Lambda$ in \eqref{Gamma2explicit}
\beq\label{Lambdaupper}
\Lambda< 5 \cdot 10^{15}{\rm GeV} \Big(\frac{m_2}{1 {\rm TeV}}\Big)^{3/2}.
\eeq
Note that the allowed range \eqref{Lambdabound}, \eqref{Lambdaupper} includes the see-saw scale $\lesssim 10^{15}$ GeV, an intriguing 
possibility in view of the leptonic-only signal in PAMELA.

We also briefly comment on the case where $\chi_1$ and/or $\chi_2$ are fermions, while leaving detailed analysis for future. In order to thermalise with the SM, the fermionic $\chi_{1,2}$ need to be either charged under the SM gauge groups or couple to a hidden sector that then mediates with the SM. The simplest case is that $\chi_{1}$ and/or $\chi_{2}$ are weak doublets so that the neutral components are massive ``dark neutrinos" (some dark multiplet must also exist to cancel anomalies). Then $\chi_{1}$ and/or $\chi_{2}$ have masses in the $10$ TeV range as in the scalar case. Again, this can be avoided by enlarging the dark matter sector. 

What would be the signatures of a 2DM model in a collider experiment? If $\chi_2$ is produced near threshold, it can decay in the detector and may be observed directly.  Let us be more specific with a few illustrative examples. If $\chi_1$ is a singlet scalar and $\chi_2$ a weak doublet fermion, then the decays $\chi_2\to \chi_1 \nu, \nu Z, \nu \gamma$ are possible. If on the other hand $\chi_1$ is a weak doublet fermion, while $\chi_2$ is a singlet scalar, the decays $\chi_2\to 2\chi_1$, $\chi_2\to \chi_1^+\chi_1^-\to X_{\rm SM}$ and $\chi_2\to \chi_1\nu, \chi_1^+l^-, l^+l^-, \bar q q$ are possible. If $\chi_{1,2}$ are both weak doublet fermions, then the possible two body decays are  $\chi_2\to \chi_1 h,\chi_1 Z$, $\chi_2\to \nu h, \nu Z$, $\chi_2\to\chi_1^+W^-, l^+W^-$. How challenging the experimental search for $\chi_2$ may be will depend on the actual masses and branching ratios.

\section{Discussion and conclusions}

The proposed  2DM  explanation of PAMELA/ATIC anomaly is in some ways reminiscent of situations already discussed in the literature, where DM is not a thermal relic, but rather originates from decays of a heavier state. Examples are for instance decays of gravitinos or weakly coupled moduli into DM \cite{Moroi:1999zb}. Also in these two cases the correlation between the annihilation cross section of DM and the relic abundance is modified and relaxed from the thermal relic relation. In this way also decaying gravitino or weakly coupled moduli decays can give
large enough "boost factors" to explain PAMELA-ATIC anomaly. An important difference with 2DM is that neither gravitino
nor moduli are thermal relics. Rather, their abundance (before the decay) is reflective of Planck scale physics. The 2DM setup 
represents the other limit, where $\chi_2$ interactions governing its abundance (before the decay) are not Planck suppressed and can possibly be probed at future colliders.  In this way 2DM model is much closer to the simple thermal relic scenario.

In conclusion, we have presented a simple 2DM model that can mimic the large boost factors $B$ needed to explain the PAMELA and balloon experiment data. Two hierarchies are needed for this proposal to work: i) the ratio of $\chi_1$ and $\chi_2$ cross sections need to be large, $\sim B$, and ii) the $\chi_2$ decay width needs to be much smaller than for normal electroweak decays.  If the PAMELA data cannot be explained using astrophysics, it may be necessary to revisit the kind of scenario outlined in this manuscript in more detail.




\begin{thebibliography}{9}
\bibitem{Adriani:2008zr}
  O.~Adriani {\it et al.},
  arXiv:0810.4995 [astro-ph];
  arXiv:0810.4994 [astro-ph].
  
\bibitem{Barwick:1997ig}
  S.~W.~Barwick {\it et al.}  [HEAT Collaboration],
  Astrophys.\ J.\  {\bf 482}, L191 (1997)
  [arXiv:astro-ph/9703192].
  
\bibitem{Aguilar:2007yf}
  M.~Aguilar {\it et al.}  [AMS-01 Collaboration],
  Phys.\ Lett.\  B {\bf 646}, 145 (2007)
  [arXiv:astro-ph/0703154].
  
\bibitem{Chen:2008dh}
  C.~R.~Chen and F.~Takahashi,
  arXiv:0810.4110 [hep-ph];
  C.~R.~Chen, F.~Takahashi and T.~T.~Yanagida,
  arXiv:0809.0792 [hep-ph];
  M.~Cirelli, R.~Franceschini and A.~Strumia,
  Nucl.\ Phys.\  B {\bf 800}, 204 (2008)
  [arXiv:0802.3378 [hep-ph]];
  J.~H.~Huh, J.~E.~Kim and B.~Kyae,
  arXiv:0809.2601 [hep-ph].
  V.~Barger, W.~Y.~Keung, D.~Marfatia and G.~Shaughnessy,
  arXiv:0809.0162 [hep-ph].
  
\bibitem{Cirelli:2008pk}
  M.~Cirelli, M.~Kadastik, M.~Raidal and A.~Strumia,
  arXiv:0809.2409 [hep-ph].
  
\bibitem{Cholis:2008hb}
  I.~Cholis, L.~Goodenough, D.~Hooper, M.~Simet and N.~Weiner,
  arXiv:0809.1683 [hep-ph];
  I.~Cholis, G.~Dobler, D.~P.~Finkbeiner, L.~Goodenough and N.~Weiner,
  arXiv:0811.3641 [astro-ph].

\bibitem{Finkbeiner:2008qu}
  D.~P.~Finkbeiner, T.~Slatyer and N.~Weiner,
  arXiv:0810.0722 [hep-ph];
  N.~Arkani-Hamed and N.~Weiner,
  arXiv:0810.0714 [hep-ph];
  N.~Arkani-Hamed, D.~P.~Finkbeiner, T.~Slatyer and N.~Weiner,
  arXiv:0810.0713 [hep-ph].
  
\bibitem{Rothstein:2009pm}
  I.~Z.~Rothstein, T.~Schwetz and J.~Zupan,
  arXiv:0903.3116 [astro-ph.HE].
  
\bibitem{Harnik:2008uu}
  R.~Harnik and G.~D.~Kribs,
  arXiv:0810.5557 [hep-ph].
  
\bibitem{Nomura:2008ru}
  Y.~Nomura and J.~Thaler,
  arXiv:0810.5397 [hep-ph].
  

  
\bibitem{Fox:2008kb}
  P.~J.~Fox and E.~Poppitz,
  arXiv:0811.0399 [hep-ph].

\bibitem{Chen:2009dm}
  F.~Chen, J.~M.~Cline and A.~R.~Frey,
  arXiv:0901.4327 [hep-ph].
  
\bibitem{Mardon:2009rc}
  J.~Mardon, Y.~Nomura, D.~Stolarski and J.~Thaler,
  arXiv:0901.2926 [hep-ph].
  
\bibitem{Meade:2009rb}
  P.~Meade, M.~Papucci and T.~Volansky,
  arXiv:0901.2925 [hep-ph].

\bibitem{Grajek:2008pg}
  P.~Grajek, G.~Kane, D.~Phalen, A.~Pierce and S.~Watson,
  arXiv:0812.4555 [hep-ph].
  
\bibitem{Bertone:2007aw}
  G.~Bertone, W.~Buchmuller, L.~Covi and A.~Ibarra,
  JCAP {\bf 0711}, 003 (2007)
  [arXiv:0709.2299 [astro-ph]].



\bibitem{Ibarra:2008jk}
  A.~Ibarra and D.~Tran,
  arXiv:0811.1555 [hep-ph].
  
\bibitem{Nardi:2008ix}
  E.~Nardi, F.~Sannino and A.~Strumia,
  JCAP {\bf 0901}, 043 (2009)
  [arXiv:0811.4153 [hep-ph]].
  
\bibitem{Arvanitaki:2008hq}
  A.~Arvanitaki, S.~Dimopoulos, S.~Dubovsky, P.~W.~Graham, R.~Harnik and S.~Rajendran,
  arXiv:0812.2075 [hep-ph].

\bibitem{Ishiwata:2009vx}
  K.~Ishiwata, S.~Matsumoto and T.~Moroi,
  arXiv:0903.0242 [hep-ph].
  
\bibitem{Shirai:2009kh}
  S.~Shirai, F.~Takahashi and T.~T.~Yanagida,
  arXiv:0902.4770 [hep-ph].

\bibitem{Cheung:2009si}
  K.~Cheung, P.~Y.~Tseng and T.~C.~Yuan,
  arXiv:0902.4035 [hep-ph].
  
\bibitem{taylor}
J. H. Taylor and D. R. Stinebring, ARA\&A {\bf 24}, 285 (1986).
 
\bibitem{backer}
D. C. Backer, {\it Pulsars} (Galactic and Extragalactic Ra- 
dio Astronomy, 1988), pp. 480â€"521. (2008). 

\bibitem{Hooper:2008kg}
  D.~Hooper, P.~Blasi and P.~D.~Serpico,
  JCAP {\bf 0901}, 025 (2009)
  [arXiv:0810.1527 [astro-ph]].

\bibitem{Yuksel:2008rf}
  H.~Yuksel, M.~D.~Kistler and T.~Stanev,
  arXiv:0810.2784 [astro-ph].

\bibitem{Profumo:2008ms}
  S.~Profumo,
  arXiv:0812.4457 [astro-ph].

\bibitem{Shaviv:2009bu}
  N.~J.~Shaviv, E.~Nakar and T.~Piran,
  arXiv:0902.0376 [astro-ph.HE].

\bibitem{Malyshev:2009tw}
  D.~Malyshev, I.~Cholis and J.~Gelfand,
  arXiv:0903.1310 [astro-ph.HE].

\bibitem{Hall:2008qu}
  J.~Hall and D.~Hooper,
  arXiv:0811.3362 [astro-ph].
  
\bibitem{Torii:2008xu}
  S.~Torii {\it et al.},
  arXiv:0809.0760 [astro-ph].

\bibitem{ATIC-2}
J. Chang, W.K.H. Schmidt, J.H. Adams et al.,
Proc. of 29th ICRC (Pune) 3, 1-4, 2005.

\bibitem{Collaboration:2008aaa}
  F.~Aharonian {\it et al.}  [H.E.S.S. Collaboration],
  Phys.\ Rev.\ Lett.\  {\bf 101}, 261104 (2008)
  [arXiv:0811.3894 [astro-ph]].
  
\bibitem{Bertone:2004pz}
  G.~Bertone, D.~Hooper and J.~Silk,
  Phys.\ Rept.\  {\bf 405}, 279 (2005)
  [arXiv:hep-ph/0404175];
  L.~Bergstrom and A.~Goobar,
{\it ``Cosmology and particle astrophysics,''}
{\it  Berlin, Germany: Springer (2004) 364 p}

\bibitem{Diemand:2008in}
  J.~Diemand, M.~Kuhlen, P.~Madau, M.~Zemp, B.~Moore, D.~Potter and J.~Stadel,
  arXiv:0805.1244 [astro-ph].
  
\bibitem{Springel:2008by}
  V.~Springel {\it et al.},
  arXiv:0809.0894 [astro-ph].
  
\bibitem{Cao:2007fy}
  Q.~H.~Cao, E.~Ma, J.~Wudka and C.~P.~Yuan,
  arXiv:0711.3881 [hep-ph].



\bibitem{Hisano:2003ec}
  J.~Hisano, S.~Matsumoto and M.~M.~Nojiri,
  Phys.\ Rev.\ Lett.\  {\bf 92}, 031303 (2004)
  [arXiv:hep-ph/0307216];
  J.~Hisano, S.~Matsumoto, M.~M.~Nojiri and O.~Saito,
  Phys.\ Rev.\  D {\bf 71}, 063528 (2005)
  [arXiv:hep-ph/0412403].
  
\bibitem{MarchRussell:2008tu}
  J.~D.~March-Russell and S.~M.~West,
  arXiv:0812.0559 [astro-ph].
  
  \bibitem{Pospelov:2007mp}
  M.~Pospelov, A.~Ritz and M.~B.~Voloshin,
  Phys.\ Lett.\  B {\bf 662}, 53 (2008)
  [arXiv:0711.4866 [hep-ph]].
  
   \bibitem{Ibe:2008ye}
  M.~Ibe, H.~Murayama and T.~T.~Yanagida,
  arXiv:0812.0072 [hep-ph].
  
\bibitem{Guo:2009aj}
  W.~L.~Guo and Y.~L.~Wu,
  arXiv:0901.1450 [hep-ph].
  
\bibitem{Goldberg:1983nd}
  H.~Goldberg,
  Phys.\ Rev.\ Lett.\  {\bf 50}, 1419 (1983).
  
\bibitem{kawasaki}
  M.~Kawasaki, K.~Kohri and T.~Moroi,
  Phys.\ Lett.\  B {\bf 625} (2005) 7
  [arXiv:astro-ph/0402490].
  
\bibitem{turner}
  M.~S.~Turner, G.~Steigman and L.~M.~Krauss,
  Phys.\ Rev.\ Lett.\  {\bf 52}, 2090 (1984).

\bibitem{gelmini}
  G.~Gelmini, D.~N.~Schramm and J.~W.~F.~Valle,
  Phys.\ Lett.\  B {\bf 146} (1984) 311.

\bibitem{dicus}
  D.~A.~Dicus and V.~L.~Teplitz,
  Phys.\ Rev.\  D {\bf 34}, 934 (1986).

\bibitem{khlopov}
A.G.Doroshkevich and M.Yu.Khlopov, 
Mon.Not.R.astr.Soc. {\bf 211} (1984) 277

\bibitem{olive}
  K.~A.~Olive, D.~Seckel and E.~Vishniac,
  Astrophys.\ J.\  {\bf 292} (1985) 1.

\bibitem{sanchez-saledo}
  F.~J.~Sanchez-Salcedo,
  Astrophys.\ J.\  {\bf 591} (2003) L107
  [arXiv:astro-ph/0305496].

\bibitem{cen}
  R.~Cen,
  Astrophys.\ J.\ {\bf 546}(2001) L77
  [arXiv:astro-ph/0005206].

\bibitem{melia}
  M.~Abdelqader and F.~Melia,
  arXiv:0806.0602 [astro-ph].

\bibitem{Moroi:1999zb}
  T.~Moroi and L.~Randall,
  Nucl.\ Phys.\  B {\bf 570}, 455 (2000)
  [arXiv:hep-ph/9906527];
  B.~S.~Acharya, P.~Kumar, K.~Bobkov, G.~Kane, J.~Shao and S.~Watson,
  JHEP {\bf 0806}, 064 (2008)
  [arXiv:0804.0863 [hep-ph]];
  G.~F.~Giudice, E.~W.~Kolb and A.~Riotto,
  Phys.\ Rev.\  D {\bf 64}, 023508 (2001)
  [arXiv:hep-ph/0005123].


\end{thebibliography}
\end{document}